\documentstyle[12pt]{article}
\newcommand{\slp}{\raise.15ex\hbox{$/$}\kern-.57em\hbox{$\partial$}}
\newcommand{\sla}{\raise.15ex\hbox{$/$}\kern-.57em\hbox{$a$}}
\newcommand{\slA}{\raise.15ex\hbox{$/$}\kern-.57em\hbox{$A$}}
\newcommand{\slb}{\raise.15ex\hbox{$/$}\kern-.57em\hbox{$b$}}

\newcommand{\be}{\begin{equation}}
\newcommand{\ee}{\end{equation}}
\newcommand{\bear}{\begin{eqnarray}}
\newcommand{\ear}{\end{eqnarray}}

\newcommand{\D}{\cal D}

\begin{document}
\begin{flushright}
HD--THEP--94--40
\end{flushright}
\quad\\
\vspace{1.8cm}
\begin{center}
{\bf\LARGE Fermionic Coset Realization}\\
\medskip
{\bf\LARGE of the Critical Ising Model}\\
\vspace{1cm}
D. C. Cabra\footnote[1]{On leave of absence from the Universidad
Nacional de La Plata, Argentina} and K. D. Rothe\\
\bigskip
Institut  f\"ur Theoretische Physik\\
Universit\"at Heidelberg\\
Philosophenweg 16, D-69120 Heidelberg\\
\vspace{1cm}
{\bf Abstract}\\
\end{center}
We obtain an explicit realization of all the
primary fields of the Ising model in terms of a conformal field
theory of constrained fermions. The four-point correlators of the
energy, order and disorder operators are explicitly calculated.

\newpage
The quantum field theoretic description of critical statistical
systems has received much attention in the past \cite{Ibb}. In
particular, the Ising model has provided a useful laboratory for
testing field-theoretical ideas and techniques. It has been known
for some time that at criticality it can be described by free massless
Majorana fermions $\psi, \bar\psi$ \cite{O, Iz}. In terms of these, the
energy operator $\epsilon(x)$ is given by the local product
$\psi\bar\psi$.
Using this representation, multicorrelators of the energy operator
 have been computed \cite{Iz}.
Some correlators involving the order operator $\sigma(x)$ and disorder operator
$\mu(x)$ have also been calculated using indirect methods. The reason is that
a local representation in terms of Majorana fermions is lacking in this case.

On the other hand, the Ising model can be described in terms of a Conformal
Quantum Field Theory (CQFT) of central charge $c=\frac{1}{2}$ \cite{BPZ},
corresponding to $k=1$ in the minimal unitary series \cite{FQS}
\be\label{1.0}
c=1-\frac{6}{(k+2)(k+3)}\ee
Using conformal methods, four-point functions involving the order and
disorder operators have been calculated \cite{BPZ} on the basis of
representation theory of the Virasoro algebra. In this approach, $\sigma
(x)$ and $\mu(x)$ are regarded as primaries of conformal dimensions
$\left(\frac{1}{16},\frac{1}{16}\right)$, with correlators satisfying
the corresponding null-vector equation.

In this letter we use the ideas of CQFT and of ref.
\cite{BRS}, \cite{nos} in order to obtain a fermionic coset realization of
the critical Ising model. In this way we arrive at a complete
description of the model, and in particular of the order and disorder
operators in terms of local products of the fundamental observables of
the theory.

The Ising model can be described by a conformal quantum field theory
of central charge $c=\frac{1}{2}$ \cite{FQS} and corresponds to a
coset model G/H with $G=SU(2)_1\times SU(2)_1$ and $H=SU(2)_2$
\cite{GKO}. In the fermionic description \cite{BRS}, \cite{nos}
this theory can be realized by making the identification
\be\label{1.1}
\frac{SU(2)_1\times
SU(2)_1}{SU(2)_2}=\frac{\frac{U(2)}{U(1)}\times\frac{U(2)}{U(1)}}
{SU(2)_2}\ee

The corresponding Lagrangian is obtained by starting from
two different kinds
of $U(2)$-fermions, and gauging the respective $U(1)$
subgroups, as well as the diagonal subgroup $SU(2)$:
\bear\label{1.2}
{\cal L}&=&\frac{1}{\sqrt2\pi}\bar\psi^i\left(({\slp}+i{\sla})
\delta_{ij}+i{\slA}_{ij}\right)\psi^j\nonumber\\
&&+\frac{1}{\sqrt2\pi}\bar\chi^i\left(({\slp}+i{\slb})
\delta_{ij}+i{\slA}_{ij}\right)\chi^j\ear
where $i,j=1,2$ and $A_\mu$ are $SU(2)$ Lie-algebra-valued fields.

We now make the change of variables \cite{Foot1}
\bear\label{1.3}
a&=&i(\bar\partial h_1)h_1^{-1},\quad\bar a=i(\partial \bar h_1)\bar h_1^{-1}
\nonumber\\
b&=&i(\bar\partial h_2)h_2^{-1},\quad\bar b=i(\partial \bar h_2)\bar h_2^{-1}
\nonumber\\
A&=&i(\bar\partial g)g^{-1},\quad\bar A=i(\partial \bar g)\bar g^{-1}
\nonumber\\
\psi_1&=&h_1g\psi_1^{(0)},\quad\psi_2^\dagger=\psi_2^{(0)\dagger}(h_1g)^{-1}
\nonumber\\
\psi_2&=&\bar h_1\bar
g\psi_2^{(0)},\quad\psi_1^\dagger=\psi_1^{(0)\dagger}(\bar
h_1\bar g)^{-1}
\nonumber\\
\chi_1&=&h_2 g\chi_1^{(0)},\quad\chi_2^\dagger=\chi_2^{(0)\dagger}(h_2g)^{-1}
\nonumber\\
\chi_2&=&\bar h_2\bar
g\chi_2^{(0)},\quad\chi_1^\dagger=\chi_1^{(0)\dagger}(\bar
h_2\bar g)^{-1}
\ear
Taking account of the Jacobians of the respective transformations
\cite{PW, Fi} (see also \cite{nos}) one arrives at a decoupled form for the
partition function associated with the Lagrangian
(\ref{1.2}):
\be\label{1.4a}
Z_{\frac{SU(2)_1\times SU(2)_1}{SU(2)_2}}=Z_F Z_BZ_{WZW}Z_{gh}\ee
where
\newpage
\bear\label{1.4b}
Z_F&=&\int{\D}\psi^{(0)}{\D}\bar\psi^{(0)}{\D}\chi^{(0)}{\D}\bar\chi^{(0)}
\exp(-\frac{1}{\pi}\int(\psi^\dagger_2\bar\partial\psi_1+\psi_1^\dagger\partial
\psi_2))\nonumber\\
&&\times \exp\
(-\frac{1}{\pi}\int(\chi^\dagger_2\bar\partial\chi_1+\chi^\dagger_1
\partial\chi_2)\ear
\be\label{1.4c}
Z_B=\int{\D}\phi_1{\D}\phi_2\exp(\frac{1}{\pi}\int\phi_1\Delta\phi_1)
\exp(\frac{1}{\pi}\int\phi_2\Delta\phi_2)\ee
\be\label{1.4d}
Z_{WZW}=\int{\D}\tilde g\exp(-kW[\tilde g])\ee
and $Z_{gh}$ is the partition function of ghosts, whose explicit form will
not be required. The fields $\phi_1$ and $\phi_2$ are defined
via
\be\label{1.5}
h_i=e^{-(\phi_i+i\eta_i)},\bar h_i=e^{(\phi_i-i\eta_i)}\ee
where $\eta_1$ and $\eta_2$ are pure gauge degrees of freedom. In arriving
at (\ref{1.4a})-(\ref{1.4d}), $\eta_1$ and $\eta_2$, as well as the
gauge-variant
combinations $\hat g=g\bar g$ have been integrated out. $W[\tilde g]$ in
(\ref{1.4d})
is the Wess-Zumino-Witten (WZW) action \cite{Wi}
\bear\label{1.6a}
W[\tilde g]&=&-\frac{1}{8\pi}\int d^2x tr(\partial_\mu\tilde
g\partial^\mu\tilde
g^{-1})\nonumber\\
&&+\frac{1}{12\pi}\int d^3y\epsilon_{ijk} tr(\tilde g^{-1}\partial_i\tilde
g\tilde g^{-1}\partial_j\tilde g\tilde g^{-1}\partial_k\tilde g)\ear
where $\tilde g$ is defined by
\be\label{1.6b}
\tilde g=g^{-1}\bar g\ee
and $k=-(2+2c_V)=-6$ is the level of the Kac Moody algebra. The negative level
WZW sector is non-unitary. However, the BRST quantization condition
connects the different sectors (decoupled at the Lagrangian level) and
ensures unitarity \cite{KO}, \cite{KS}. The central charge obtained by
adding the individual contributions $c_F=4$, $c_B=2$, $c_{WZW}=9/2$ and $c_{gh}
=-10$ is equal to $c=1/2$ in accordance with the central charge of the
Ising model.

The observables of the theory defined by the Lagrangian (\ref{1.2}) will be
constructed from the gauge-invariant operator products
\newpage
\bear\label{1.7}
\hat\psi^i(x)&=&e^{-i\int^\infty_xdz_\mu a_\mu}\left(Pe^{-i\int^\infty_xdz^\mu
A_\mu}\right)_{ij}\psi^j(x)\nonumber\\
\hat\chi^i(x)&=&e^{-i\int^\infty_xdz_\mu b_\mu}\left(Pe^{-i\int^\infty_xdz^\mu
A_\mu}\right)_{ij}\chi^j(x)\ear
As we now show, all the primaries of the Ising model can be obtained as local
products of the operators (\ref{1.7}) in the isospin zero sector. In this
sector the non-abelian Schwinger line integrals associated with the diagonal
$SU(2)$-gauge group will not contribute.

We begin by identifying the energy operator $\epsilon$ and order operator
$\sigma$
with isospin neutral local operator products of the fundamental fields
(\ref{1.7}):
\bear\label{1.8}
\epsilon&=&\left((\hat\psi_2^\dagger\hat\chi_1+\hat\chi_2^\dagger\hat\psi_1)
(\hat\chi_1^\dagger\hat\psi_2+\hat\psi_1^\dagger
\hat\chi_2)\right) \nonumber\\
&=&\left(\psi_2^{(0)\dagger}\chi_1^{(0)}\right)\left(\chi_1^{(0)\dagger}
\psi_2^{(0)}\right)
e^{2\phi_1}e^{-2\phi_2}
\nonumber\\
&&+(\psi_2^{(0)\dagger}
\chi_1^{(0)})(\psi_1^{(0)\dagger}\chi_2^{(0)})e^{\varphi_1-\bar\varphi_1}
e^{-(\varphi_1-\bar\varphi_2)}
\nonumber\\
&&+(\psi\leftrightarrow \chi,\quad \varphi_1\leftrightarrow
\varphi_2,\quad\bar\varphi_1\leftrightarrow\bar\varphi_2)\ear
\bear\label{1.9}
\sigma&=&\left(\hat\psi^\dagger\hat\psi+\hat\chi^\dagger
\hat\chi\right)\nonumber\\
&=&(\psi_2^{(0)\dagger}\tilde g\psi_2^{(0)})
e^{2\phi_1}+(\psi_1^{(0)\dagger}\tilde
g^{-1}\psi^{(0)}_1)e^{-2\phi_1}\nonumber\\
&&+(\psi\to\chi, \phi_1\to\phi_2)\ear
\bear\label{1.10}
&&\varphi_i=\phi_i+i\int^\infty_xdz_\mu\epsilon_{\mu\nu}\partial_\nu\phi_i
\nonumber\\
&&\bar\varphi_i=\phi_i-i\int^\infty_xdz_\mu\epsilon_{\mu\nu}\partial_\nu\phi_i
\ear
are the holomorphic and antiholomorphic components of the fields $\phi_i$ and
where we have identified the normal-ordered products $:gg^{-1}:$ and
$:\bar g\bar g^{-1}:$ with the identity operator. The relative signs
in (\ref{1.8}) and (\ref{1.9}) have been chosen such as to conform to the
operator product expansions in the literature \cite{FSZ}. The conformal
dimension of the fields can be read off from the two-point
functions \cite{Foot2}
\bear\label{1.11a}
&&\langle\psi_1^{(0)^{i_1}}(1)\psi_2^{(0)^{i_2}\dagger}(2)\rangle=\frac{1}{2}
\frac{\delta^{i_1i_2}}{z_{12}}\nonumber\\
&&\langle\psi_2^{(0)^{j_1}}(1)\psi_1^{(0)^{j_2}\dagger}(2)\rangle=\frac{1}{2}
\frac{\delta^{j_1j_2}}{\bar z_{12}}\nonumber\\
&&\nonumber\\
&&\langle e^{\alpha\varphi(1)}e^{-\alpha\varphi(2)}\rangle=(\mu
z_{12})^{\frac{\alpha^2}{8}}\nonumber\\
&&\langle e^{\alpha\bar\varphi(1)}e^{-\alpha\bar\varphi(2)}\rangle=(\mu \bar
z_{12})^{\frac{\alpha^2}{8}}\nonumber\\
&&\nonumber\\
&&\langle g^{i_1j_1}(1)g^{-1j_2i_2}(2)\rangle=|\mu z_{12}|^{3/4}\delta
^{i_1i_2}\delta^{j_1j_2}\ear
where $z_{ij}=z_i-z_j$, $\bar z_{ij}=\bar z_i-\bar z_j$.
The conformal dimensions of $\epsilon$ and $\sigma$ are
obtained by adding the dimensions of the components making up
(\ref{1.8}) and (\ref{1.9}), and are found to be $h_\epsilon=\bar h_\epsilon=
\frac{1}{2}$ and $h_\sigma=\bar h_\sigma=\frac{1}{16}$, respectively.

{}From (\ref{1.8}) we see that all multipoint correlation functions
of $\epsilon$ can be calculated in terms of the free-fermion and
vertex two-point functions. Special attention has to be payed to the fact
that with the definitions (\ref{1.10}), we have
\be\label{1.11b}
\langle e^{\alpha\varphi(1)}e^{\beta\bar\varphi(2)}\rangle =e^{i\alpha\beta
\frac{\pi}{4}}\ee
For the four-point function of $\epsilon$ one thus obtains
\be\label{1.12}
\langle\epsilon(1)\epsilon(2)\epsilon(3)\epsilon(4)\rangle
=\frac{1}{|z_{12}z_{34}|^2}\frac{|1-x+x^2|}{|1-x|^2}=
\left|P_f\left(\frac{1}{z_{ij}}\right)\right|^2\ee
where
\be\label{1.12a}
x=\frac{z_{12} z_{34}}{z_{14}z_{32}},\quad \bar x=\frac{\bar z_{12}\bar
z_{34}}{\bar z_{14}\bar z_{32}}.\ee
Expression (18) agrees with the expected result \cite{Iz}.

The computation of the four-point function of the $\sigma$-field
requires the knowledge of the  four-point functions
\newpage
\bear\label{1.13}
&&G(1,2,3,4):=\langle tr(\tilde g(1)\tilde g^{-1}(2))tr(\tilde g(3)\tilde
g^{-1}(4))\rangle\nonumber\\
&&\hat G(1,2,3,4):=\langle tr(\tilde g(1)\tilde g^{-1}(2)\tilde g(3)\tilde
g^{-1}(4))\rangle\ear
of the level (-6) WZW field, and those obtained by permutation of
the arguments.
These functions can be directly extracted from the results of ref. \cite{KZ}.
However, for their computation it is advantageous to make use of the
equivalence
between the $SU(2)_2$ WZW theory and the fermionic coset
model $U(4)/(SU(2)_2\times U(1))$\cite{NS}. For the
particular arrangement of the arguments (1,2,4,3) one arrives
at the particularly simple expressions
\bear\label{1.14a}
&&G(1,2,4,3)=8|z_{14}z_{23}|^{3/4}\frac{|x|}{|x(1-x)|^{1/4}}(f_1(x)f_1(\bar
x)+f_2(x)f_2(\bar x))\nonumber\\
&&\hat G(1,2,4,3)=-8|z_{14}z_{23}|^{3/4}\frac{\sqrt{x(1-\bar
x)}}{|x(1-x)|^{1/4}}(f_1(x)f_2(\bar x)-f_2(x)f_1(\bar x))\ear
where
\be\label{1.14c}
f_1(x)=\sqrt{1+\sqrt x},\quad f_2(x)=\sqrt{1-\sqrt x}.\ee
{}From here the corresponding expressions for permutations of the arguments
are easily computed.
Note that the functions $f_i$  are the independent solutions of the
hypergeometric equation arising from the null-vector equation for a
primary field of conformal dimension 1/16 \cite{BPZ}.

One can show after a lengthy, but straightforward calculation that the
four-point
correlator of the order operator (\ref{1.9}) is given by the expression
\bear\label{1.15}
&&\langle\sigma(1)\sigma(2)\sigma(3)\sigma(4)\rangle=\nonumber\\
&&\frac{1}{|z_{14}z_{23}|^{1/4}}\frac{1}{|x(1-x)|^{1/4}}(\sqrt{1+\sqrt x}
\sqrt{1+\sqrt{\bar x}}+\sqrt{1-\sqrt x}\sqrt{1-\sqrt{\bar x}}),\ear
which agrees with that obtained by BPZ \cite{BPZ} using
general conformal arguments. (In order to compare with the BPZ result we note
that expression (\ref{1.15}) is invariant under any permutation of the indices,
and in particular under $z_3\leftrightarrow z_4$, implying $x\rightarrow
y=\frac{x}{x-1}$.)

The disorder operator should satisfy the equal-time dual algebra \cite{Schroer}
\be\label{1.18}
\sigma(1)\mu(2)=e^{i\pi \theta(x_1-x_2)}\mu(2)\sigma(1)\ee
This leads us to make the following ansatz in terms of the
gauge-invariant fermion fields (\ref{1.7})
\bear\label{1.16}
\mu&=&\hat\psi^\dagger\hat\chi+\hat\chi^\dagger
\hat\psi\nonumber\\&=&(\psi_2^{\dagger(0)}\tilde
g\chi_2^{(0)})e^{\varphi_1+\bar\varphi_2}+(\chi^{\dagger(0)}_1\tilde g^{-1}
\psi_1^{(0)})e^{-(\varphi_1+\bar\varphi_2)}\nonumber\\
&&+(\psi\leftrightarrow\chi,\varphi_1\leftrightarrow\varphi_2,
\bar\varphi_1\leftrightarrow\bar\varphi_2).\ear
Using (15) one checks that this operator has dimension 1/16, as required.
With the aid of
the (euclidean) equal-time commutators:
\bear\label{1.17}
&&[\varphi(1),\varphi(2)]_{ET}=-\frac{i\pi}{2} {\rm sgn}(x_{12})\nonumber\\
&&[\varphi(1),\bar\varphi(2)]_{ET}=\frac{i\pi}{2}{\rm sgn}(x_{12})\nonumber\\
&&[\varphi(1),\bar\varphi(2)]_{ET}=\frac{i\pi}{2}\ear
one furthermore checks that the order-disorder algebra (\ref{1.18})
is satisfied.

The evaluation of the four-point correlator of the disorder operator
parallels the one for the order operator, the result being again given by
the r.h.s. of eq. (\ref{1.15}), as it should be \cite{KC}.

The calculation of the mixed order-disorder correlator $\langle
\sigma(1)\mu(2)\sigma(3)\mu(4)\rangle$ proceeds along  similar
lines, and only involves
the WZW four-point function in the combinations (\ref{1.13}). Also due
account has to be taken of the phases arising from (\ref{1.11b}). A
straightforward calculation shows that the four-point functions arising
from the individual terms in the expressions (\ref{1.9})
and (\ref{1.16}) for $\sigma(x)$ and $\mu(x)$, respectively, though
seemingly different at first sight, have all the same form,
and in fact contribute with the same weight to the $\sigma-\mu$ correlator.
The result of the calculation is most conveniently expressed in terms of the
variable $y=\frac{x}{x-1}$, and takes the form
\be\label{1.19}
\langle\sigma(1)\mu(2)\sigma(3)\mu(4)=
\frac{1}{|z_{13}z_{24}|^{1/4}}\frac{1}{|y(1-y)|^{1/4}}
\left(\sqrt{1+\sqrt y}\sqrt{1-\sqrt{\bar y}}-\sqrt{1-\sqrt y}\sqrt{1+
\sqrt{\bar y}}\right)\ee
which agrees with the result obtained by BPZ on the basis of general conformal
considerations, giving further support to our ans\"atze.

To complete our discussion of the Ising model, we give a realization
of the Onsager fermions $\psi(x)$ and $\bar\psi(x)$ \cite{O}, \cite{Iz}
in the fermionic coset framework. They are naturally identified with the
gauge-invariant composites
\bear\label{1.20}
\psi&=&\hat\psi^\dagger_2\hat\chi_1+\hat\chi^\dagger_2\hat\psi_1\nonumber\\
\bar\psi&=&\hat\psi^\dagger_1\hat\chi_2+\hat\chi^\dagger_1\hat\psi_2\ear
of dimensions $\left(\frac{1}{2},0\right)$ and $\left(0,\frac{1}{2}\right)$,
respectively.

Recalling (\ref{1.8}) we see that the assignment (\ref{1.20})
is in agreement with the usual representation of the energy operators in
terms of Majorana fermions \cite{Iz}. Note that they have the expected property
$\psi^{\dagger}=\bar\psi$.

Our explicit calculations lend support to the general validity of our
fermionic coset realization of {\cal all the primary operators} of the critical
Ising model in terms of local products of BRST-invariant fermionic fields.
In the framework of Majorana fermions, this
is not possible. Similarly, the bosonic coset description \cite{GK} does not
lend itself to a natural identification of all the primaries of the Ising
model.

In our fermionic coset formulation of the Ising model the identification
 became possible at the price
of allowing for an underlying non-abelian structure realized by
WZW fields, free massless fermions and bosons. In particular the primaries of
the Ising
model have been represented in terms of local BRST invariant products
of free Dirac fermions, vertex operators
of free massless bosons, and negative-level WZW fields. Thus the calculation of
multipoint correlators of the primaries is only restricted by our lack
of knowledge of $n$-point correlators of WZW fields, with $n>4$.
It interesting to note, that according to (\ref{1.9}) the application of an
external magnetic field in the Ising model correspond to the addition of a mass
term in the fermionic coset formulation. On the other hand, an off-critical
perturbation linear in the energy operator is described acccording to
(\ref{1.8})
by a quartic self-interaction of fermions.

Our construction for the Ising model can be generalized to other
statistical mo\-dels corresponding to the minimal unitary  series,
with the Ising model corresponding to $k=1$. Work in this direction will be
reported elsewhere.

Our construction may also prove useful for studying perturbations
of minimal models about the critical point. The same applies to the study of
statistical systems in the presence of line defects \cite{Be},\cite{McC.P}.

\bigskip
\noindent{\bf Acknowledgement:} One of the authors (D.C.C.) would like
to thank the Commission of the European Community for the ``Marie Curie
Fellowship'', which made this collaboration possible. We also thank J\"urgen
Fuchs for a useful discussion.


\bigskip
\begin{thebibliography}{12}
\bibitem{Ibb} C. Itzykson, H. Saleur, and J.-B. Zuber,
Editors, ``Conformal Invariance and Applications to
Statistical Mechanics'', World Scientific 1988.
\bibitem{O} L. Onsager, Phys. Rev. {\bf 65}, 117 (1944).
\bibitem{Iz} M. Bander and C. Itzykson, Phys. Rev. {\bf D15}, 463 (1977);\\
C. Itzykson, and J. M. Drouffe, {\it Statistical Field Theory}, Vol. 1,2,
Cambridge Univ. Press 1989.
\bibitem{BPZ} A. A. Belavin, A. M. Polyakov, and A. B. Zamolodchikov,
Nucl. Phys. {\bf B241}, 333 (1981).
\bibitem{FQS} D. Friedan, Z. Qiu, and S. Shenker, Phys. Rev. Lett.
{\bf 52}, 1575 (1984).
\bibitem{BRS} E. Bardacki, E. Rabinovici, and B. Saring, Nucl. Phys. {\bf
B229},
151 (1988).
\bibitem{nos} D. Cabra, E. Moreno, and C. von Reichenbach, Int. J.
Mod. Phys. {\bf A5}, 2313 (1990).
\bibitem{GKO} P. Goddard, A. Kent, and D. Olive, Int. J. Mod. Phys.
{\bf A1}, 303 (1986).
\bibitem{Foot1} Our notation is
$a=\frac{1}{\sqrt2}(a_1+ia_2),\ z=\frac{1}{\sqrt2}(x+iy),\ \partial=
\frac{\partial}{\partial z}$.
\bibitem{PW} A. Polyakov and P. Wiegmann, Phys. Lett. {\bf 131B}, 121 (1983)
\bibitem{Fi} R. E. Gamboa Saravi, F. A. Schaposnik, and J. E. Solomin, Nucl.
Phys. {\bf B185}, 238 (1981).
\bibitem{Wi} E. Witten, Comm. Math. Phys. {\bf 92}, 455 (1984).
\bibitem{KO} T. Kugo and I. Ojima, Suppl. Progr. Theor. Phys. {\bf 66}, 1
(1979).
\bibitem{KS} D. Karabali, Q. Han Park, H. Schnitzer, and Z. Yang,
Phys. Lett. {\bf 216B}, 307 (1989).
\bibitem{FSZ} P. Di Francesco, H. Saleur, and D. B. Zuber, Nucl. Phys.
{\bf B29} [FS20], 527  (1987).
\bibitem{Foot2} In what follows the arbitrary scale parameter $\mu$
in (16) will be conveniently chosen such as to conform with the
results in the literature.
\bibitem{KZ} V. G. Knizhnik and A. B. Zamolodchikov, Nucl. Phys. {\bf B247}, 83
(1984).
\bibitem{NS} S. Naculich and H. Schnitzer, Nucl. Phys. {\bf B333}, 583 (1990);
{\bf B347}, 687 (1990).
\bibitem{Schroer} B. Schroer and T. T. Truong, Nucl. Phys. {\bf B154}, 125
(1979).
\bibitem{KC} L. P. Kadanoff and H. Ceva, Phys. Rev. {\bf B3}, 3918 (1971).
\bibitem{GK} K. Gawedzki and A. Kupiainen, Nucl. Phys. {\bf B320}, 624 (1989).

\bibitem{Be} R. Bariev, Sov. Phys. JETP {\bf 50}, 613 (1979).
\bibitem{McC.P}B. McCoy and J. Perk, Phys. Rev. Lett. {\bf 44}, 840 (1980).


\bibitem{CR} D. C. Cabra and K. D. Rothe, in preparation.
\end{thebibliography}
\end{document}